\begin{filecontents*}{}
\documentclass[twocolumn]{svjour3}          
\smartqed  
\usepackage{graphicx}
\usepackage{subcaption}
\usepackage{amsmath}
\usepackage{nccmath}
\usepackage{amsfonts}
\graphicspath{{./figures/}}
\DeclareGraphicsExtensions{.png}
\usepackage{multirow}
\usepackage[table]{xcolor}
\usepackage{textcomp}
\usepackage{url}
\usepackage{mathptmx}      
\begin{document}

\title{Interpreting Imagined Speech Waves with Machine Learning techniques
}


\author{Abhiram Singh         \and
        Ashwin Gumaste 
}


\institute{All authors are at Computer Science and Engineering Department, Indian Institute of Technology Bombay, India-400076 \\
              \email{abhiram25.1990@gmail.com, ashwin@ashwin.name}  \\
}


\maketitle

\begin{abstract}
This work explores the possibility of decoding Imagined Speech (IS) signals which can be used to create a new design of Human-Computer Interface (HCI). 
Since the underlying process generating EEG signals is unknown, various feature extraction methods, along with different neural network (NN) models, are used to approximate data distribution and classify IS signals. 
Based on the experimental results, feed-forward NN model with ensemble and covariance matrix transformed features showed the highest performance in comparison to other existing methods. 
For comparison, three publicly available datasets were used. We report a mean classification accuracy of 80\% between rest and imagined state, 96\% and 80\% for decoding long and short words on two datasets. 
These results show that it is possible to differentiate brain signals (generated during rest state) from the IS brain signals. 
Based on the experimental results, we suggest that the word length and complexity can be used to decode IS signals with high accuracy, and a BCI system can be designed with IS signals for computer interaction. 
These ideas, and results give direction for the development of a commercial level IS based BCI system, which can be used for human-computer interaction in daily life.
\keywords{Brain-Computer Interface \and Imagined Speech \and Neural Networks \and Tangent Space}
\end{abstract}

\section{Introduction}\label{sec:intro}
Brain-Machine Interface (BMI) is a collection of software (for analyzing cognitive tasks) and hardware components (used to capture brain signals). Nowadays, BMI research is gaining momentum to diagnose brain disease, its possible use in human-computer interface (HCI) devices, and to study human behavior. When used as an HCI device, a BCI system can evolve with computer interaction technologies such as a keyboard, touch screen, or mouse. There exist many BMI systems for Human-Computer Interaction (HCI) \cite{wolpaw} such as motor imaginary or P300.

A known fact about the human brain is that it generates electrical signals while performing different activities. One such class of brain signals is Imagined speech (IS) \cite{Hickok2007TheCO}. In IS condition, a person speaks in mind without moving any articulators. Note that the silent speech signals are different from IS, in which a user thinks about the articulator's movement for pronunciation of words. Previous studies suggest that the source of IS signals are Broca's and Wernicke's area \cite{Sahin2009SequentialPO}, whereas the motor cortex of the brain is considered a primary source of silent speech signals. 

With improvement in the technology, there exists different techniques to record electrical signals of the brain. Electroencephalography (EEG) \cite{Michal2002EEG} is one such non-invasive system, that involves placing electrodes over the scalp. EEG electrodes capture voltage differences generated due to ions movement inside the brain region. The voltage differences are captured and stored for a time duration to generate an EEG signal. The number of EEG electrodes (1-256) are selected based on the experiment requirements. 

The objective of this work is to explore different techniques that improve decoding capability of IS signals. A motivation to work on IS signals is to reduce the training time of participants and provide a more comfortable procedure for HCI than the motor imaginary tasks. Also, IS based BMI system offers a natural way towards HCI, which can lead to improved user experience in computer interaction. One assumption for the IS system to work is that the data is not fully corrupted. Therefore it is possible to extract IS speech related information. Subjects participating in IS experiments follow specific guidelines to make this assumption feasible.

The work presented in this paper identifies a classification model and useful discriminative features to improve decoding performance on IS signals. Based on the experimental results on different datasets, tangent space (TS) \cite{Barachant2012Riemann} turns out to be the most discriminative input feature to an artificial feed forward neural networks (ANN) \cite{Goodfellow2016DL} model. Our approach of TS+ANN improves the classification accuracy from  72.6\% to 79.3\%, 49.3\% to 60.16\%, and 49.2\% to 57.83\%  on one long vs. short word, three short words and three vowels classification tasks respectively.  

The following sections of this paper are as follows. Section 2 introduces the problem statement, and our contribution, along with an overview of existing methods to decode IS signals. Section 3 describes various classification models and feature extraction methods. Section 4 provides dataset detail and shows the result on different classification models in combination with appropriate feature extraction methods. Finally, section 5 discusses a few points for decoding IS signals and concluding remarks.

\section{Problem Statement and Literature Review}\label{sec:ps}
In this section, we formally describe problem statement, our contributions and prior related work.  

\subsection{Problem Statement and Contribution}
\textit{Problem Statement}: For a given EEG signal, our goal is to identify signal category (imagined speech or other kind of brain signal). If signal belongs to the IS category, then we want to 1) decode word category; 2) actual imagined word.

\textit{Contribution}: 1) We evaluate the generalization performance of neural networks (NN) on raw IS EEG signals. We show experimentally that for the given raw EEG signals of IS, the NN models fail to generalize. This leads us to the further exploration of various feature extraction techniques that are likely to improve NN performance. Main result: we identify tangent space \cite{BARACHANT2013172} as the most useful discriminative feature and feed-forward NN as the most successful classification model for decoding IS. Results confirm that feature engineering indeed improved classifier performance. NN classifier performance was further improved by using an ensemble technique. Using experimental results, we also show that the presence of high-frequency components in an IS signals is required for enhancing the classifier performance. We compare and show results on three publicly available IS datasets and our approach outperformed existing approaches on different IS tasks.
2) We show that the feature extraction method used to decode IS signals is also capable of discriminating participant's rest state EEG signals from the IS signals.

\subsection{Related work} Zhao et al. \cite{Zhao} suggested support vector machines (SVM) with two different kernels and deep-belief networks (DBN) as classification models. The authors in \cite{Zhao} show results for each subject imagining some word or a phoneme. However, there are only about 10-15 trials per subject per word/phoneme. Hence, the classification result shown is on the limited amount of data and, therefore, does not have significant statistical importance. Rather than showing results for individual subjects, we have combined data of different subjects to obtain about 170 trials for each class (word/phoneme). This data was then used to compare the performance of different learning models. Suitability of the provided dataset lies in the fact that the data is of good spatial and temporal resolution.   

Nguyen et al. \cite{Nguyen2018IS} created three IS tasks comprising 2 long words, 3 short words and 3 vowels. Authors suggested covariance matrix based features (projected in Tangent Space (TS)) along with Relevance Vector Machine (RVM) \cite{Psorakis2010RVM} to decode IS signals. Results (in \cite{Nguyen2018IS}) show the classification accuracy of 49\%, 50\% and 66\% for vowels, short words, and long words respectively. We significantly improve these results by using TS features with principal component analysis (PCA) \cite{Jolliffe2014pca} to reduce feature dimensionality as well as use NN models for classification and ensemble techniques to improve model performance.

In a similar context, Pradeep et al. \cite{kumar2018envisioned} proposed envisioned speech recognition using statistical features of the EEG signal and random forest as the classifier.
Tianwei et al. \cite{shi2019feature} proposed a wavelet-reconstruction approach for extracting useful frequency band from EEG signals and a convolution neural network based classification model for decoding the motor imagery based EEG signals.
Menezes et al. \cite{menezes2017towards} proposed emotion recognition based on the EEG signals using support vector machine and random forest classifiers with input as the statistical features and band power from different frequency bands.

\section{Proposed Approach for Imagined Speech Decoding}\label{sec:pa}
Our proposed approach for decoding the IS signal is summarized in the following steps. First, we attempt to decode raw EEG signals by feeding the signal directly as an input to different NN models. Raw EEG input will enable automatic feature extraction from the NN models leading to the learning of target class distribution. Second, we start exploring different feature extraction methods. For each feature extraction method, we select a classification model to represent temporal or spatial dependency among features. Third, we compare the performance of feature extraction, and associated classification models with the best IS decoding model. The above steps give us a comparative analysis of different features and classification models in the domain of IS signals. Finally, we compare our proposed approach with existing approaches for decoding IS signals.

\subsection{Background on the Classification Models}\label{sec:models}
We now briefly describe the three NN models that are used for feature transformation and estimation of input conditioned target class distribution. The parameters of all models are updated using the Adam optimization algorithm \cite{Kingma2014adam}. Gradient computation was performed with respect to the cross-entropy loss, and gradient propagation was performed using back-propagation. 

The artificial neural network (ANN) model \cite{Goodfellow2016DL} performs two steps in an iterative fashion: 1. linearly combine the output of previous hidden layer; 2. apply a non-linear activation function to generate the desired output:
\begin{equation}\label{eq:ann}
a^l= g^l(M^la^{l-1})
\end{equation}
Where, vector $a^{l-1} \in \mathbb{R}^n$ represents activations from the previous layer $l-1$, $a^l \in \mathbb{R}^m$ represents activations of layer $l$, weight matrix $M^l \in \mathbb{R}^{(m,n)}$ contains tunable parameters between layer $l-1$ and $l$, and $g^l$ is the ReLU activation function at the hidden layer $l$. 

To capture dependency between inputs, some nodes in the network can have self-loops. This allows nodes to propagate the information across multiple inputs. The following relations govern the recurrent neural network (RNN) \cite{Goodfellow2016DL}: 
\begin{equation}\label{eq:gnn}
h^{(i)} = g(W_{hh} h^{(i-1)} + W_{hi} f^{(i)});\; t_{pred}^{(i)} = g(W_{oh} h^{(i)})
\end{equation}
Where $f^{(i)} \in \mathbb{R}^n $ is the input,  $t_{pred}^{(i)} \in \mathbb{R}^m $ is output, $h^{(i)} \in \mathbb{R}^k $ is hidden layer, $ W_{hi}\in \mathbb{R}^{(k,n)} $ is the weight matrix between input to hidden layer, $ W_{hh}\in \mathbb{R}^{(k,k)} $ is the weight matrix connecting hidden layer nodes to other hidden layer nodes, $ W_{oh}\in \mathbb{R}^{(m,k)} $ represents a kernel matrix from hidden to output layer, $g$ is usually a non-linear activation function.

In convolution neural networks (CNN) \cite{Goodfellow2016DL} \cite{MatthewCNN}, a neuron is activated only for certain regions of a visual area known as a receptive field. Since neurons share parameters in one layer of the network and the receptive field dimension is much lower in comparison to an input dimension, the number of training parameters in CNN is much less than the number of parameters in the ANN for the same input dimension. A convolution operation is defined as the inner product between the weights of a layer (kernel/filter) and input neurons in the receptive field. This inner product is computed iteratively by shifting the window over the uncovered neurons. Some non-linear function and pooling layer usually follow the convolution layer. Feature map dimension after a convolution operation can be given as $1 +(n – k + 2p)/s$. Where the input is of size $(n,n)$, while the kernel is of size $(k,k)$. The stride length is $s$, and padding size is $p$. We tried with different filter sizes to capture time and temporal dependency among EEG channels. To do so, we used max-pooling, and varied the number of convolution and pooling layers, used ReLU and leaky-ReLU activation functions to improve overall model performance.

\subsection{Approach}
This section describes the various approaches that we took in decoding the IS signals. 

\textit{Raw EEG signal + ANN:} 
We start with raw EEG signals and use three different ANN models, (which are described in section \ref{sec:models} for feature extraction and estimation of conditional class distribution. Since ANN takes a vector as its input, the signal of each channel is appended along the time dimension to form a vector. For example, if a single EEG trial is of dimension $(c,t)$ where $c$ and $t$ represents the number of channels and samples in the trial, then a column vector of dimension $(t\times c,1)$ is formed. A trial is a sequence of signal samples across different channels that corresponds to some cognitive activity. The limitation of creating a vector is the high dimensionality of the input space in comparison to the available number of trials. The ANN model has a large number of parameters to tune and due to the limited size of the dataset, overfitting occurs in the early stage of learning. 

\textit{Raw EEG signal + RNN:} The next choice was the LSTM model \cite{lstm}, \cite{Goodfellow2016DL}, which can capture time-dependency among input features. However,  the high-sampling frequency of the EEG signal generated a long series of time-dependent signals. Despite down-sampling, it was not possible to reduce the sampling rate below 100 as most of the energy of the EEG signal is contained in the frequency band 0-50 Hz. For a single EEG trial of $(c,t)$ dimension, a $c$-dimensional vector is given as input to LSTM for $t$ time-steps. Out of the $t$ outputs generated from LSTM, only the last output was used to identify the target class and measure cross-entropy loss.

\textit{Raw EEG signal + CNN:} Next, we choose CNN because of its automatic feature extraction capabilities and successful application on a wide variety of EEG tasks \cite{Lawhern}. The two dimensional EEG input of the form $(c,t)$ is given as input directly to the CNN to automatically extract features from raw EEG data and determine the appropriate target class.

\textit{Fourier Coefficients + ANN:} We can also analyze the signal in another basis. The change of basis provides a different view and reveals the hidden properties of the signal. A change of basis might also serve as a dimension reduction technique, as the number of useful features can be reduced in another dimension. With this intuition, we start to explore a few signal transformation techniques. 
We can remove the time dependency by applying a signal transformation technique. In this case, the signal analysis was done in the Fourier domain (see \cite{FT} for details). The Fourier transformation (FT) provides a change of basis. Fourier coefficients are computed as:
\begin{equation}\label{eq:fft}
Y[w] = \sum_{n=-\infty}^{\infty} y[n] e^{-iwn}
\end{equation}
where, $y[n]$ is the discrete time signal, $e^{-iwn}$ is the complex exponential at frequency, $w$ and $Y[w]$ is the Fourier coefficient corresponding to the frequency $w$.
Fourier coefficients representing frequency up to 40 Hz were combined to form a vector. This Fourier coefficient vector was given as an input feature to the ANN model. FT was performed using Python Numpy library \cite{Numpy}. 

\textit{Spectrogram + CNN:} By tracking frequency shifts with time, it might be possible to identify the effects of IS condition on the brain signals. To obtain both time and frequency representation of a signal, a spectrogram (see \cite{stft}) is used. The spectrogram is just stacking of frequency features over a time-axis, calculated using Short Time Fourier Transforms (STFT). The spectrogram was computed using the Python Scipy library \cite{Scipy}. For each channel of a trial, the spectrogram of dimension $[f,t]$ was calculated. By combining the spectrogram of each channel,  we obtained data of dimension $[f,t,c]$, which was given as an input to a CNN, to track frequency shift with time along each channel. Where $f$ is the number of frequency bins, $t$ is the number of time bins, and $c$ is the number of channels in an EEG trial.

\textit{Independent Components + CNN:} The signal captured at EEG electrodes can be seen as a linear combination of signals generated due to a group of neuron activations and possibly due to some noise such as muscle artifact, eye movement, eye blinking, or environmental noise. During experiments, artifacts generated from the body are not present in all trials of a given target class, or even if they are present, then their presence will also be realized in other target classes. So the information provided by such artifacts in the experimental environment will not be useful in discriminating target classes for the given IS based EEG trials. Independent component analysis (ICA) \cite{Ica} estimates the sources by assuming that source signals are linearly combined to generate observed signals, and source generating processes are independent of each other. The fastICA package \cite{Ica}, \cite{sklearn} was used to find independent components of brain signals, and these new components were given as $[c,t]$ - dimensional input to CNN. 

\textit{Spatial Patterns + ANN:} Common Spatial Pattern (CSP) \cite{Blankertz2008} transforms the signals such that the variance of the signal is different for different classes. We can use the variance of each channel as an input feature to the classification model. Due to the use of target class information in features transformation step, this method is expected to improve classifier performance in comparison to the unsupervised features transformation techniques. CSP performs linear transformation as follows:
\begin{equation}\label{eq:csp1}
y^{new} = Ly
\end{equation}
Where $y \in \mathbb{R}^n, y^{new} \in \mathbb{R}^n$ and transformation matrix $L \in \mathbb{R}^{(n,n)}$.
In the above, the classification performance gets improved by using $y^{new}$ in comparison to $y$. For a two-class problem, CSP can be seen as an optimization problem:
\begin{equation}\label{eq:csp2}
l^* = argmax_{l\in \mathbb{R}^n} \Bigg(\frac{l^T A_{y|c_1} l}{l^T A_{y|c_2} l} \Bigg)
\end{equation}
Where $l \in \mathbb{R}^n $ is the transformation vector, $A_{y|c_1}$ and $A_{y|c_2} \in \mathbb{R}^{(n,n)}$ are the covariance matrix of data belonging to class 1 and 2, $l^*$ is the optimized transformation vector. For dimension reduction, transformation matrix $ L \in \mathbb{R}^{(j,n)} $ with $j < n$ can be used. Solution to optimization problem in equation \ref{eq:csp2} can be obtained by solving the following problem:
\begin{equation}\label{eq:csp3}
A_{y|c_1} l = \omega A_{y|c_2} l
\end{equation}
where $\omega$ is the eigenvalue and $l$ is corresponding eigenvector. After transformation of EEG signals according to equation \ref{eq:csp1} (CSP package \cite{mne}), the variance of each channel is calculated to form a $c$-dimensional vector and given as an input feature to the ANN.

\textit{Tangent Space + ANN:} For a given EEG trial, it is possible to measure the spatial dependency between EEG channels by computing the covariance matrix. The covariance matrices must be represented in a vector form to apply various dimension reduction techniques. However, the projection must preserve the discriminative information of the target class. To this end, covariance matrics are projected to the tangent space (TS) \cite{Barachant2012Riemann} involving the following matrix operations:
\begin{equation}\label{tsEq}
\centering
\begin{split}
P_i = C_m^{1/2}logm(C_m^{-1/2}C_iC_m^{-1/2})C_m^{1/2} \\
logm(N) = AB'A^{-1}, \; B'[i,i] = log(B[i,i])
\end{split}
\end{equation}
Where $C_i$ is an input covariance matrix, $C_m$ represents the mean of covariance matrices, $P_i$ is the required projection, $N$ is a diagonalizable matrix, $N^{-1}$ is the matrix inverse, and $ABA^{-1}$ is the decomposition of the matrix $N$. Same decomposition $ABA^{-1}$ of matrix $N$ can be used to compute $N^{-1/2}$.
The final transformation step flattens the matrix $P_i$ to obtain a vector representation. The dimension of flattened vector is reduced using principal component analysis (PCA) \cite{Jolliffe2014pca} and then given as an input to the ANN. Computation of tangent vectors was performed using the Pyriemann library \cite{Pyriemann}. 

\textit{Bagging + ANN:} We used the bootstrap aggregation (Bagging) classifier \cite{Breiman1996bag} with ANN as its base classifier. Bagging classifier trains multiple base classifiers on a small random subsets of the dataset. The result of each base classifier is combined by averaging, which provides a single output of the bagging classifier. Bagging improves mean classification accuracy and reduces the variance of the base classifier. We used a bagging classifier for the ANN classifier only. Training of CNN and LSTM is computationally intensive, and hence bagging was not used with these classifiers.

\subsection{Data Visualization} 
For data visualization in two dimensions, we use PCA and tSNE \cite{tsne}. The objective function of PCA is,
\begin{equation}\label{eq:pca}
\begin{split}
max_{u\in\mathbb{R}^n}\; u^TAu \\
subject\; to\; \|u\|_2^2 = 1
\end{split}
\end{equation}
Here, $A$ is the data covariance matrix defined as $\frac{1}{m} \sum_{i=1}^m x^{(i)} {x^{(i)}}^T $ and $ u, x^{(i)} \in \mathbb{R}^n $. The optimization problem (defined in equation \ref{eq:pca}) can be solved by obtaining a solution to the Eigenvalue problem $ Au = \omega u $.

The t-distributed stochastic neighbor embedding (t-SNE) computes probabilities by using the distance between points as a measure. The computed probabilities define distribution in a higher and lower dimension. The difference between these two distributions is minimized by using Kullback-Leibler (KL) divergence as a cost function. KL divergence is defined as,
\begin{equation}\label{eq:tsne}
KL(P,Q) = \sum_{i \neq j}p_{ij}log\frac{p_{ij}}{q_{ij}}
\end{equation}
Where $P,Q$ are two distributions defined in higher and lower dimensions, respectively. t-SNE implementation [22] was used for dimension reduction.

\section{Dataset Details and Results}
Now we discuss the datasets in detail, explain various pre-processing steps, show the performance of our proposed approach on these datasets, and finally compare obtained results with existing approaches. 
\subsection{Dataset details and Pre-processing}
We use three publicly available datasets given by Zhao et al. \cite{Zhao}, Nguyen et al. \cite{Nguyen2018IS} and Coretto et al. \cite{Coretto}, which we shall term as \textit{dataset1}, \textit{dataset2} and \textit{dataset3} respectively. \textit{Dataset1} contains 11 categories \textit{(iy, uw, piy, tiy, diy, m, n, pat, pot, knew, gnaw)} from 13 subjects. We call EEG data corresponding to an output label, a trial. We reject a few trials because these contain relatively high or low values, possibly due to some error in the experiments. The remaining 1735 trials were low-pass filtered with an upper cut-off frequency of 40Hz. After that, each trial was down-sampled to 128Hz from 1000Hz. We restrict the number of samples in each trial to 619 to remove variation across trials. Finally, trials corresponding to each output class were appended sequentially to form a 3D matrix of dimension $[1735, 62, 619]$ and targets of each category as one hot 11-dimensional vector. 

\textit{Dataset2} contains eight categories \textit{(a, i, u, in, out, up, independent, cooperate)} with three subcategories having three vowels, three short words and two long words. Each category contains 100 trials per subject. For vowels and short words, the trial duration was of 1 second, and for long words, the trial duration was 1.4 seconds. The dataset was down-sampled to 256Hz. In a trial, subjects performed three repetitive thinking of the same word or vowel. Therefore each trial provides three matrices of $[c, t]$ dimension.  For long words matrix dimension is $[60, 360]$  and for vowels and short words matrix dimension is $[60, 256]$. Therefore, we have $[600, 60, 360]$ or $[900, 60, 256]$ dimensional data matrix for each subject and 2 or 3-dimensional one-hot vector representation as target classes.

\textit{Dataset3} contains 11 categories \textit{(a, e, i, o, u, up, down, left, right, backward, forward)} with data collected from 15 subjects. Each vowel or word has approximately 50 trials repeated in random order by each subject. However, only six electrodes are used in the experiment and the sampling frequency is 1024Hz. In each trial, the imagined speech duration was set to 4 seconds following the rest state condition, which was for another 4 seconds. Data of each subject was stored in the format $[N, C, T]$ where $N$ is the number of trials for any subject, $C$ is the number of channels, and $T$ is the number of samples in that trial. In our case, if a subject is having exactly 50 trials for each category then, the input is a $[550, 6, 4096]$ dimensional matrix and 11 or 2-dimensional one-hot vector (depending on the decoding of all categories simultaneously or decoding vowels vs. words) as the target. Data in the above format was taken as processed input to different methods, and then we performed feature transformation and decoding (as described in section \ref{sec:pa}).

\subsection{Results}
In this section, different feature extraction methods and classification models are evaluated for decoding IS signals.

\subsubsection{Performance metric} 
Classification accuracy (CA) is used as a metric to compare the performance of different approaches. CA calculates the fraction between the number of correct predictions vs. total predictions. Accuracy results are reported using the 10-fold cross-validation scheme. Each fold involves data separation (using the stratified sampling) into a train and test set. Stratified sampling maintains the sample proportion of each class in the train and test sets.

\subsubsection{Results on \textit{dataset1}} We follow an incremental approach in which, based on initial stage results, we select only top-performing feature extraction methods and related models. We start evaluating the performance of different feature extraction and classification models with \textit{dataset1}. The results are shown in Figure \ref{dataset11} where each column shows accuracy after decoding 11 categories of the IS based EEG signal. From Figure 1, it is clear that CSP and TS feature extraction methods outperform other methods. Bagging also slightly helps in the performance improvement. Other feature extraction methods (described in section \ref{sec:pa}) do not generalize to the classification task, and hence we report further results on \textit{dataset2} and \textit{dataset3} using CSP and TS based methods. Using TS and CSP, we also checked individual performance for seven phonemes and four words, as shown in Figure \ref{dataset12}. Here, both methods gave slightly improved results when compared to the chance level accuracy. Chance level accuracy in seven phonemes and four words task is 14.28\% and 25\%, respectively.

\begin{figure}[t]
    \centering
    \includegraphics[width=0.99\linewidth, height=5cm]{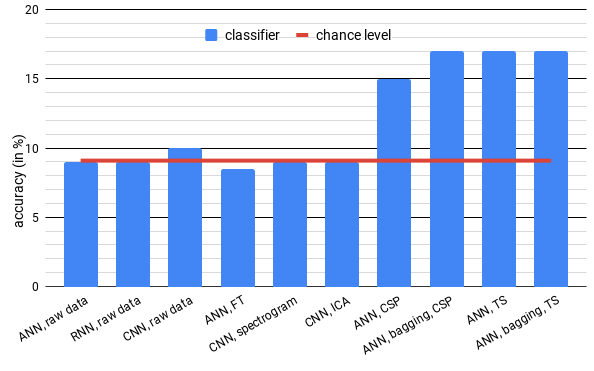}
    \caption{11 class classification on \textit{dataset1}.}
    \label{dataset11}
\end{figure}

Subsequently, we checked whether it is possible to discriminate between words based on their length. We combined trials of words and phonemes in two different groups and applied CSP and TS methods with ANN. In this case, we obtained high accuracy (Figure \ref{dataset12}). CSP+ANN showed 94\% accuracy, and TS+ANN showed 96\% accuracy. Bagging slightly improved performance in both cases. Chance level accuracy in the word-vs-phoneme classification task is 50\%. 

It is pertinent to note that in Figure \ref{dataset12}, we compared results by combining the data of all the subjects. This requirement came from the fact that there are only about 10-15 trials per subject per class. Model learning and performance comparison cannot be done on very few training and testing examples. Combining data of different subjects might hurt the model performance, but it increased the number of training examples to about ten times (for each target class). Increased number of training samples helped us in identifying the best feature extraction method and model for the IS recognition task.

\textit{Rest vs. IS state}: To develop a real-time IS based BCI system, it is necessary to distinguish between IS and rest-state conditions. For classifying IS signal from rest-state brain signal, we extracted rest-state EEG signals from the available dataset and applied the same pre-processing steps as for the case of imagined speech signals. Then trials corresponding to imagined speech signals and rest state signals were divided into two different groups to form a binary classification problem. We applied CSP+ANN and TS+ANN models to solve binary classification problem. Using CSP features, 71\% accuracy, and using TS features, 79\% accuracy is obtained (Figure \ref{dataset12}). These results confirm that IS EEG signals carry a lot of discriminative information from brain signals generated during the rest-state.  
\begin{figure}[t]
    \centering
    \includegraphics[width=0.95\linewidth, height=5cm]{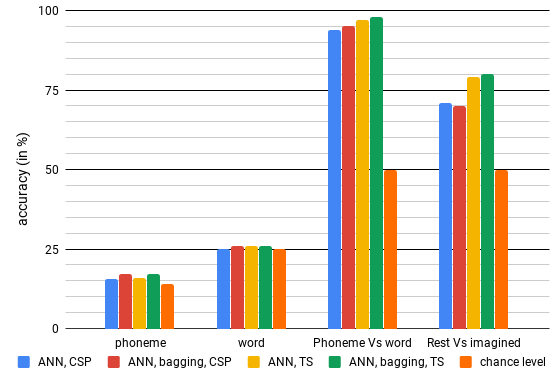}
    \caption{Classification on four different tasks of \textit{dataset1}.}
    \label{dataset12}
\end{figure}

\begin{figure*}[t]
    \centering
    \includegraphics[width=0.9\linewidth, height=7cm]{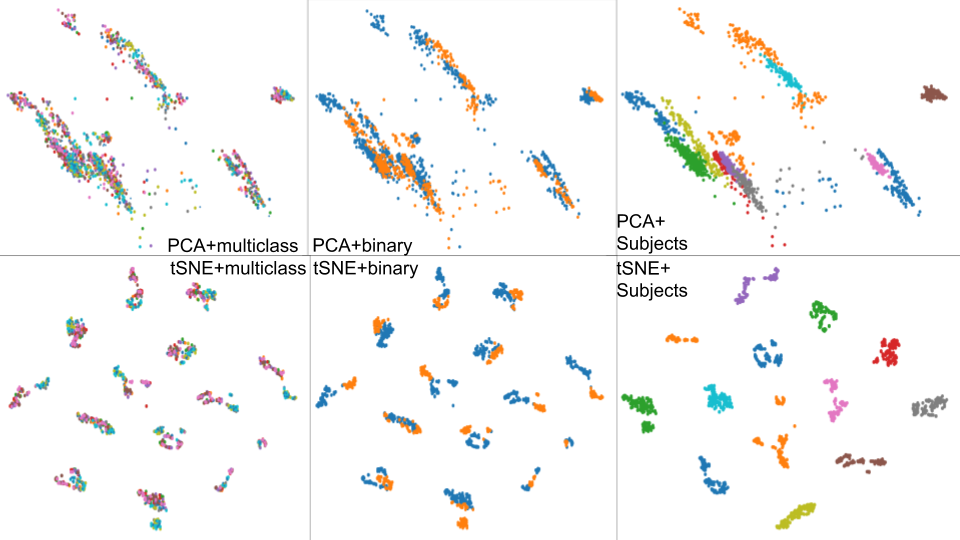}
    \caption{TS in the lower dimension using PCA and tSNE for \textit{dataset1}. Two figures on the left show data points for 11 target classes, and two figures in the middle show binary classification between words and phonemes. The rightmost figures show data points for each subject. Top figures are obtained using PCA and bottom are obtained using tSNE. Each color in the left and central figure represents a target class of that point, and each color in the rightmost figures belongs to different subjects. The rightmost figure using tSNE, shows one cluster per subject.}
    \label{dataset13}
\end{figure*}

\textit{Dataset1 visualization}: Results on \textit{dataset1} show very low accuracy on the multiclass problem and high accuracy on binary classification. This point is worth investigation, and we understand this behavior by visualization of points in lower dimensions (see Figure \ref{dataset13}). For visualization, we projected the TS features to 2 dimensions using PCA and tSNE. Projected points are plotted with different colors, with each color representing the target class for that point. 

As we can see from Figure \ref{dataset13}, the points for the multiclass problem are mixed up heavily. PCA tends to form 5 to 6 clusters and each cluster has data from multiple classes. Hence it becomes difficult to find a clear separation boundary. In higher dimensions, points may be separated. However, classification results show that even in a higher dimension, a clear separation cannot be obtained with a good generalization capability. When observing data using tSNE, about 13-14 well-separated clusters are formed. However, data-points of different classes continue to be present within each cluster, which makes it difficult for a classifier to separate points. For binary classification using PCA, we can see that data of two classes is overlapping with a slight shift of points belonging to different classes. Using tSNE for binary classification between words and phonemes, the separation between points belonging to each cluster is more visible. The visualization for binary classification becomes more evident in the higher dimensions as the classifier can create a separation boundary with good generalization performance on the test set. 

By observing the data in a lower dimension using the t-SNE approach, we can see some clusters. Counting of these clusters was approximately equal to the number of participants in the experiment. The next obvious question was to check if these clusters are indeed representing different participants. To this end, data of different participants was appended sequentially. TS based features were extracted, and then PCA and t-SNE on these features were applied and plotted for each subject (rightmost part of Fig \ref{dataset13}). PCA and t-SNE both show clusters for each participant, but the division is visible only in the latter approach. The visualization for each subject clearly shows a possible application of the IS based EEG signals in the domain of biometric-based human authentication.

\subsubsection{Results on \textit{dataset2}} To further check the robustness of the proposed approach, the performance of these methods was compared on \textit{dataset2}. Due to the availability of data-points, we show results for each subject, by training and testing on individual subjects’ data. Figure \ref{dataset2} shows results for 3 short words, 2 long words, 3 vowels, and 1 short vs. long word, respectively. A comparison of CSP+ANN and TS+ANN is made with results provided Nguyen et al. \cite{Nguyen2018IS} who proposed tangent space (TS) with Relevance Vector Machine (RVM). Our findings on \textit{dataset1} clearly show that Bagging gave improved results. To this end, we used CSP+ANN and TS+ANN with Bagging for comparing results with the TS+RVM approach. We observe that the TS+RVM approach dominates over the CSP approach, but our proposed approach of using TS+ANN+ Bagging outperformed TS+RVM approach on 3 vowels (Figure \ref{dataset2vowel}) and 3 short words (Figure \ref{dataset2short}) classification tasks. Further, our approach performed equally well or better for many subjects on long vs. short words (Figure \ref{dataset2shortlong}) and 2 long words (Figure \ref{dataset2long}) classification task. Note that chance level accuracy for 3 vowels and 3 short words classification is 33.33\%, and for 2 long words and one short vs. long word classification is 50\%.

\begin{figure*}[t]
    \centering
    \begin{subfigure}[t]{0.25\textwidth}
        \includegraphics[width=0.99\textwidth,height=3.5cm]{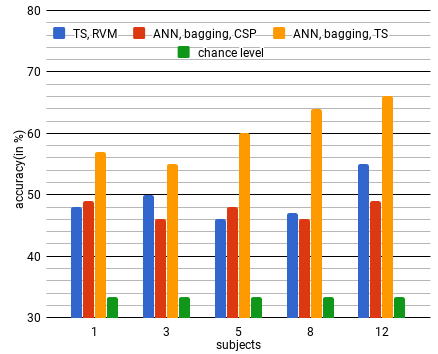}
        \centering
        \caption{3 short words.}
        \label{dataset2short}
    \end{subfigure}%
    \begin{subfigure}[t]{0.25\textwidth}
        \includegraphics[width=0.99\textwidth,height=3.5cm]{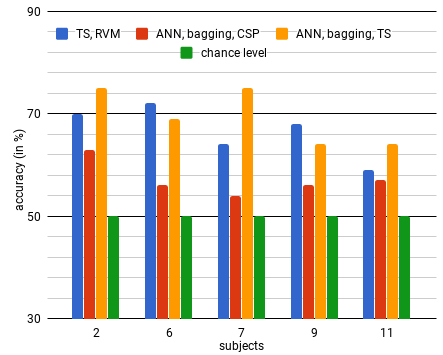}
        \centering
        \caption{2 long words.}
        \label{dataset2long}
    \end{subfigure}%
    \begin{subfigure}[t]{0.25\textwidth}
        \includegraphics[width=0.99\textwidth,height=3.5cm]{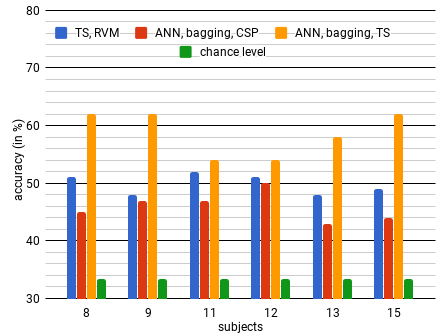}
        \centering
        \caption{3 vowels.}
        \label{dataset2vowel}
    \end{subfigure}%
    \begin{subfigure}[t]{0.25\textwidth}
        \includegraphics[width=0.99\textwidth,height=3.5cm]{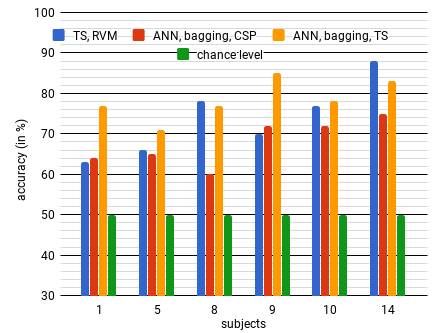}
        \centering
        \caption{1 short vs long word.}
        \label{dataset2shortlong}
    \end{subfigure}%

    \caption{Evaluation of the proposed approach (CSP and TS with ANN+bagging classifiers) on four different tasks of \textit{dataset2} and comparison with TS+RVM approach.}
    \label{dataset2}
\end{figure*}

\textit{Comparison with existing approaches}: 
Now we compare four existing approaches of IS decoding with our proposed approach (TS+ANN+bagging). First, we compare our method with the approach suggested by Tomioka et al. \cite{TOMIOKA2010415}, which applies CSP to the input data, calculates the log of the variance of each channel and uses linear discriminant analysis (LDA) as a classifier. The second comparison is made with Dasalla et al. \cite{Dasalla2009}, in which authors suggest using CSP based feature transformation technique with support vector machine (SVM) as a classifier. The third comparison is made with Min et al. \cite{Min2016}. This approach uses mean, variance, standard deviation, and skewness as the input features to the extreme learning machine (ELM) used as a classifier. The final comparison is made with Nguyen et al. \cite{Nguyen2018IS} approach of using TS as input features to ELM. We have already shown the results of Nguyen et al. \cite{Nguyen2018IS} method of using TS as an input feature to the RVM classifier (Figures \ref{dataset2}). Figure \ref{dataset2comp} shows the performance of four existing approaches with our proposed approach on one long vs. one short word, 3 short words, and 3 vowels classification tasks, respectively. Our proposed approach of using TS+ANN+bagging outperformed all four existing approaches. The performance improvement is observed across all the subjects performing three different IS tasks.

\begin{figure*}[t]
    \centering
    \begin{subfigure}[t]{0.33\textwidth}
        \includegraphics[width=0.99\textwidth,height=3.5cm]{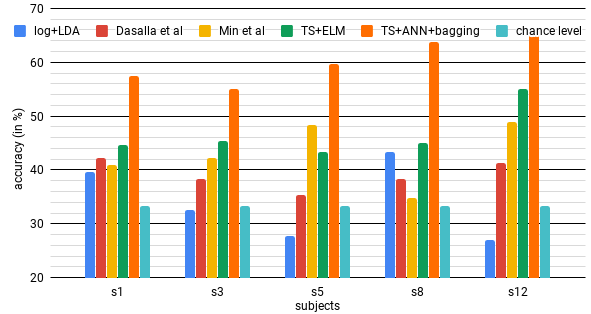}
        \centering
        \caption{3 short words.}
        \label{dataset2shortcomp}
    \end{subfigure}%
    \begin{subfigure}[t]{0.33\textwidth}
        \includegraphics[width=0.99\textwidth,height=3.5cm]{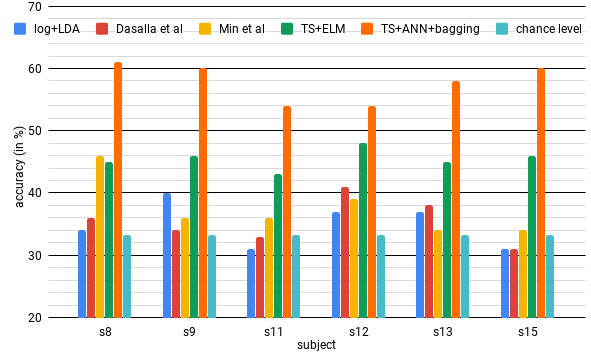}
        \centering
        \caption{3 vowels.}
        \label{dataset2vowelcomp}
    \end{subfigure}%
    \begin{subfigure}[t]{0.33\textwidth}
        \includegraphics[width=0.99\textwidth,height=3.5cm]{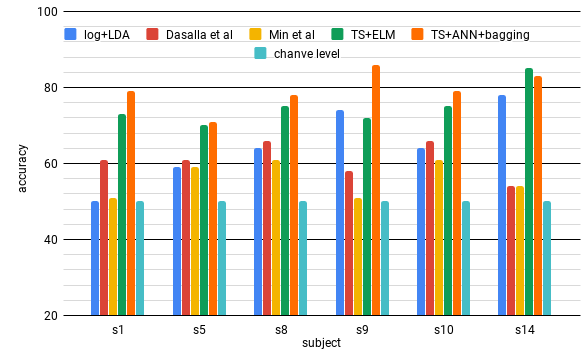}
        \centering
        \caption{1 short vs long word.}
        \label{dataset2shortlongcomp}
    \end{subfigure}%

    \caption{Comparison of the proposed approach (ts+ANN+begging) with existing approaches on three IS tasks of \textit{dataset2}.}
    \label{dataset2comp}
\end{figure*}

From Figure \ref{dataset2comp}, we observe the variation in the performance of each approach across different subjects. 
A more interpretable way for the comparison is to compute a single result for each approach. To this end, the accuracy of all subject within a task is averaged to obtain one performance measure. Table \ref{meanstd4IStasksSubAvgTab} shows the results for each task.

\begin{table}[h]
\caption{Mean classification accuracy ($\mu$) and standard deviation ($\sigma$). Results are rounded to the nearest integer. N.A. stands for results not available in \cite{Nguyen2018IS}.}
\newcolumntype{P}[1]{>{\centering\arraybackslash}m{#1}}
\begin{tabular}{|P{2cm}|P{0.4cm}|P{0.8cm}|P{0.7cm}|P{0.7cm}|P{0.7cm}|}
 \hline
 \rowcolor{lightgray}
 \multicolumn{2}{|c|}{Task} & Vowels & Short words & Short vs. Long & Long words \\
 \hline
 \multirow{2}{4em}{lda+csp} & $\mu$ & 35 & 34 & 65 & \multirow{2}{3em}{N.A.} \\
 & $\sigma$ & 4 & 7 & 10 & \\
 \hline
 \multirow{2}{4em}{elm+ts} & $\mu$ & 45 & 47 & 75 & \multirow{2}{3em}{N.A.} \\
 & $\sigma$ & 2 & 5 & 5 & \\
 \hline
 \multirow{2}{4em}{svm+csp} & $\mu$ & 35 & 39 & 61 & \multirow{2}{3em}{N.A.} \\
 & $\sigma$ & 4 & 3 & 5 & \\
 \hline
 \multirow{2}{4em}{rvm+ts} & $\mu$ & 49 & 50 & 73 & 66 \\
 & $\sigma$ & 2 & 3 & 9 & 5 \\
 \hline
 \multirow{2}{4em}{elm+statistical features} & $\mu$ & 37 & 43 & 56 & \multirow{2}{3em}{N.A.} \\
 & $\sigma$ & 5 & 6 & 5 & \\
 \hline
 \multirow{2}{4em}{ann+ts} & $\mu$ & \textbf{58} & \textbf{60} & \textbf{79} & \textbf{69} \\
 & $\sigma$ & 3 & 4 & 5 & 5 \\
 \hline
 \end{tabular}
\label{meanstd4IStasksSubAvgTab}
\end{table}

It is evident from Table \ref{meanstd4IStasksSubAvgTab} that the proposed approach (ts+ann) obtains the highest classification accuracy across different classification tasks.
The low deviation of the proposed approach implies that ANN model with begging reduces the variability across subjects.
A desired approach should provide high mean accuracy with low variance, where mean and variance is calculated across all subjects.
Our approach appears to reach that point.
Existing approaches for decoding IS signal either obtain high accuracy and high variance or low accuracy and low variance across different subjects. 
This implies that existing approaches exibits one of the following two cases: 1. Decode IS signals of only a few subjects with high accuracy, thereby showing high variance. 2. Unable to extract IS based discriminative information, thereby resulting low accuracy and small deviation.

Experimental results (presented in this paper) show the generalization capability of the ANN model in decoding IS signals when appropriate input features are provided. Note that the classification accuracy of words in the \textit{dataset1} is much lower than \textit{dataset2}. High accuracy on \textit{dataset2} suggests that the word complexity and length provide useful discriminative information in IS word decoding tasks.

\textit{High-frequency component (HFC) analysis}: From the work of Emily et al. \cite{Emily2014}, Martin et al. \cite{Martin2016} and Herff et al. \cite{Herff2015} (in which authors utilized electrocortigography (ECoG) to measure IS based brain signals), it is quite evident that authors mainly used signals with high-frequency components and thereby obtained good decoding results. However, due to the invasive nature of ECoG, the approach suggested by authors is limited to the medical patients. Despite limitation, the ECoG provides enough motivation to check whether we can use high-frequency components (HFCs) of EEG signals to improve decoding capability of a classifier. The common reason for not trying HFCs is because the signal strength drops rapidly with an increase in HFCs, and the SNR becomes low. Hence, it becomes difficult to separate signal from noise.  

To study the importance of HFCs above 50Hz in IS based EEG signals, the data belonging to two long words as a classification task was band-pass filtered between 80 and 125Hz. After that, the TS based feature extraction technique was applied with ANN and bagging. Post-classification, the results were then compared with an unfiltered signal and a 40Hz low-pass filtered signal belonging to the same two long words task. Figure \ref{dataset2highpass} shows slight performance improvement as compared to the unfiltered signal. The lowest performance was obtained when classifying the 40Hz low-pass filtered signal. We can see up to 5\% difference in accuracy between the HFCs signal and low-pass signal for up to 40Hz. These results were consistent for all the subjects.  

\begin{figure}[t]
    \centering
    \includegraphics[width=0.99\linewidth, height=5cm]{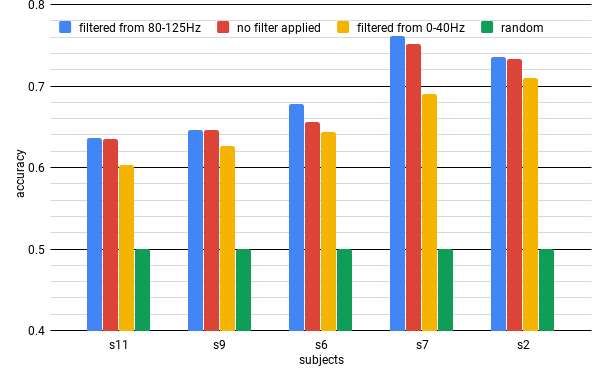}
    \caption{Using the band-pass filtered signal for two long words classification task on \textit{dataset2}.}
    \label{dataset2highpass}
\end{figure}

The usual trend in EEG signal processing is to apply a low-pass filter up to 40Hz and then extract information from the filtered signal. This result clearly shows that even if the signal amplitude is much lower for HFCs, the discriminative information continues to be present in those components. Hence low-pass filtering of IS based EEG signals is not always the right choice.

\textit{Hyper-parameter details}: Now, we provide hyperparameter details on a 2 long words classification task applied on \textit{dataset2} (Table \ref{hyperparameter}). We used ANN with one hidden layer and first row of the Table \ref{hyperparameter} (named neurons) provides the number of neurons in the hidden layer. Bagging requires that the number of classifiers to be specified (estimators row in Table \ref{hyperparameter}). We have 60 EEG channels in \textit{dataset2}, implying that the length of the tangent vector was 1830. For dimension reduction, we used PCA. In Table \ref{hyperparameter}, the row named dimensions specifies the number of components retained after applying PCA.

\begin{table}[t]
\caption{Hyper-parameter setting and model accuracy for two long words classification task on \textit{dataset2}.}
\centering
\begin{tabular}{|c|c|c|c|c|c|}
 \hline
 \rowcolor{lightgray}
 Subjects & s11 & s9 & s6 & s7 & s2 \\
 \hline
 Neurons & 100 & 200 & 50 & 150 & 50 \\
 \hline
 Estimators & 100 & 200 & 100 & 10 & 50\\
 \hline
 Dimensions & 20 & 70 & 20 & 20 & 20 \\
 \hline
 \end{tabular}
\label{hyperparameter}
\end{table}

\subsubsection{Results on \textit{dataset3}} 
Finally, we apply our approach of TS+ANN on \textit{dataset3} \cite{Coretto} in which six electrodes were used to capture EEG signals during the IS recognition task.

We show results for eleven and binary classification in Table \ref{dataset3tab}. The length of the feature vector after transformation is 21. So, after applying TS, dimension reduction using PCA is not required. Therefore, 21 dimensional feature vector is directly given as an input to the bagging classifier, which uses ANN as a base classifier. 

Due to the imbalance of data in \textit{dataset3} for the binary classification task, along with classification accuracy we also report the area under the curve (AUC) in the result. AUC measures an area under the receiver operating characteristic (ROC) curve. The ROC curve is plotted by measuring the true-positive rate (TPR) and false-positive rate (FPR) of the classifier for different values of threshold. An AUC of 0.5 shows random performance from the classifier. On the other hand, an AUC value of 1 implies that the classifier can correctly separate the classes. Hence, we want a classifier with a higher AUC value. Table \ref{dataset3tab} shows classification accuracy and AUC for different subjects. The chance level accuracy for binary classification (between vowels vs. words) task is 0.5 and that for 11 class classification is 0.0909.

\begin{table}[t]
\caption{Subjects performance on vowels vs words (binary) and 11 (multi) class classification task on \textit{dataset3}. accBinary and aucBinary represents model accuracy and AUC for the binary classification task, accMulti and aucMulti represents accuracy and AUC for the multiclass classification task.}
\centering
\begin{tabular}{|c|c|c|c|c|c|}
 \hline
 \rowcolor{lightgray}
 Subjects & s13 & s3 & s11 & s14 & s1 \\
 \hline
 accBinary & 0.669 & 0.678 & 0.685 & 0.685 & 0.626 \\
 \hline
 aucBinary & 0.662 & 0.671 & 0.677 & 0.681 & 0.625\\
 \hline
 accMulti & 0.117 & 0.147 & 0.114 & 0.114 & 0.125 \\
 \hline
 aucMulti & 0.536 & 0.582 & 0.552 & 0.551 & 0.555 \\
 \hline
 \rowcolor{lightgray}
 Subjects & s6 & s5 & s7 & s15 & s8 \\
 \hline
 accBinary & 0.719 & 0.591 & 0.572 & 0.632 & 0.766 \\
 \hline
 aucBinary & 0.721 & 0.586 & 0.555 & 0.627 & 0.76\\
 \hline
 accMulti & 0.109 & 0.103 & 0.097 & 0.144 & 0.127 \\
 \hline
 aucMulti & 0.565 & 0.505 & 0.488 & 0.559 & 0.624 \\
 \hline
 \rowcolor{lightgray}
 Subjects & s4 & s12 & s10 & s9 & s2 \\
 \hline
 accBinary & 0.572 & 0.559 & 0.65 & 0.651 & 0.661 \\
 \hline
 aucBinary & 0.561 & 0.549 & 0.643 & 0.649 & 0.605\\
 \hline
 accMulti & 0.113 & 0.135 & 0.12 & 0.111 & 0.118 \\
 \hline
 aucMulti & 0.545 & 0.542 & 0.54 & 0.563 & 0.543 \\
 \hline
\end{tabular}
\label{dataset3tab}
\end{table}

We observe low accuracy both in multiclass and binary classification task. The reason for low accuracy is the fewer number of electrodes, which reduces the presence of useful discriminative information in the data. Also, the signal was low-pass filtered up to 40Hz, which drops the classifier performance (as shown in the result of \textit{dataset2}). Another reason is the choice of words based on their meaning. We have seen from results on \textit{dataset1} and \textit{dataset2} that the word complexity is useful discriminative information, that helps in improving classifier performance.
It is clear that words in \textit{dataset3} are not complex (in terms of their speech representation), and hence word decoding is more difficult even in the binary classification task.

\section{Discussion and Conclusion}
In this section, we provide some reasoning behind the variation of model performance across three datasets and after that we provide concluding remarks.

\subsection{Discussion}
We observe the highest multiclass classification performance on \textit{dataset2}, highest binary classification on \textit{dataset1} and least performance on \textit{dataset3}. On \textit{dataset2}, multiclass classification accuracy is high (as compared to the other two datasets) because words or vowels differ in their speech signal representation. Therefore, a process generating the imagined speech signals possibly activates the neurons at different time intervals to generate activation patterns. Different activation patterns lead to the discriminative IS based EEG signals. Generally, long words are more difficult in the imagined speech pronunciation in comparison to the vowels and short words. 
The additive complexity contained in the IS signals provides more discriminative information, thereby improving the classifier performance.  

For \textit{dataset1}, we observe similar phonemes and words in terms of speech pronunciation. Hence, during the imagination of similar kind of words, internal brain representation might be similar, thereby reducing the discriminative information present in the input features. Therefore, we observe low performance using multiclass classification tasks. However, the same phonemes and words can be grouped to provide substantial discriminative information giving good accuracy for binary classification tasks between phonemes and words. 

High accuracy on the binary classification task shows that the brain has a similar representation for words that sound similar to each other. The result also holds for binary classification of vowels and words. The accuracy is lower in \textit{dataset3} (in comparison to \textit{dataset1} and \textit{dataset2}) primarily due to using very few electrodes and low-pass filtered signal. The performance of classifier for the multi-class classification task is is also low because of the similar reasons. One additional point for low accuracy in \textit{dataset3} is that the words are selected based on meaning rather than difference in the sound, length, and complexity.

\subsection{Conclusion}
The work presented in this paper shows a technique to design and develop an imagined speech based BMI system with the help of machine learning techniques.
In doing so, we explored various feature engineering methods and different neural network models to understand the decoding capability of IS signals. 
Subsequently, we proposed an approach involving covariance matrix, tangent space, principal component analysis, and artificial neural network with begging as a classifier.
The proposed approach outperformed existing approaches when applied to three publicly available datasets. 
We show that IS signals contain some discriminative information that can be used when differentiating IS signals from other brain signals (such as rest state brain signals).
We also suggest that the length and complexity of a word are a useful criterion while discriminating against a group of words. 
The future work will be dedicated on creating a machine learning model that can directly decode raw EEG signals and recover from noisy signals.

\bibliographystyle{spmpsci}      
\bibliography{reference}   

\end{document}